\documentstyle[12pt]{article}

\begin{document}
\begin{titlepage}
\begin{center}
{\Large\bf On the Sensitivity of the Polarized Parton Densities\\
\vskip 0.5cm
to Flavour SU(3) Symmetry Breaking}
\end{center}
\vskip 2cm
\begin{center}
{\bf Elliot Leader}\\
{\it Birkbeck College, University of London\\
Malet Street, London WC1E 7HX, England\\
E-mail: e.leader@bbk.ac.uk}\\
\vskip 0.5cm
{\bf Aleksander V. Sidorov}\\
{\it Bogoliubov Theoretical Laboratory\\
Joint Institute for Nuclear Research\\
141980 Dubna, Russia\\
E-mail: sidorov@thsun1.jinr.ru}
\vskip 0.5cm
{\bf Dimiter B. Stamenov \\
{\it Institute for Nuclear Research and Nuclear Energy\\
Bulgarian Academy of Sciences\\
blvd. Tsarigradsko Chaussee 72, Sofia 1784, Bulgaria\\
E-mail:stamenov@inrne.bas.bg }}
\end{center}

\vskip 0.3cm
\begin{abstract}

Motivated by the possibility that it may be misleading to assume SU(3) 
symmetry in the analysis of hyperon $\beta$-decays, we study the sensitivity 
of the polarized parton densities, derived from the world data on inclusive 
polarized deep inelastic lepton-nucleon scattering, to variation of the value 
of the nonsinglet axial charge $a_8$ from its SU(3) value of 3F-D.\\


\end{abstract}

\end{titlepage}

\newpage
\setcounter{page}{1}
Our most precise knowledge of the internal partonic structure of
the nucleon has come from decades of experiments on unpolarized
Deep Inelastic Scattering (DIS) of leptons on nucleons. More
recently there has been a dramatic improvement in the quality of
the data on {\it polarized} DIS and consequently an impressive 
growth in the precision of our knowledge of the polarized parton
densities in the nucleon. However, it will be a long time before
the polarized data, for the moment limited to neutral current
reactions, can match the unpolarized data in volume and accuracy.
As a consequence, almost all analyses of the polarized parton
densities supplement the DIS (large $Q^2$) data with
information stemming from low-$Q^2$ weak interaction reactions.
More specifically, it is conventional to use the values of
$G_A/G_V$ from neutron $\beta$-decay, and 3F-D from hyperon
$\beta$-decays to help to pin down the values of the first
moments of certain combinations of parton densities:
\begin{equation}
a_3 = (\Delta u +\Delta\bar{u})(Q^2) - (\Delta d + \Delta\bar{d})(Q^2)
={G_A\over G_V}(n\rightarrow p)\equiv g_A~,
\label{a3ga}
\end{equation}
\begin{equation} 
a_8=(\Delta u +\Delta\bar{u})(Q^2) + (\Delta d + \Delta\bar{d})(Q^2) 
-2(\Delta s+\Delta\bar{s})(Q^2)=\rm {3F-D}~,
\label{3FD}
\end{equation}
where $a_3$ and $a_8$ are the nonsinglet axial charges corresponding 
to the $\rm 3^{rd}$ and $\rm 8^{th}$ components of the axial vector 
Cabibbo current. In (\ref{3FD}) F and D are the SU(3) parameters involved 
in the matrix elements describing hyperon $\beta$-decays.

The Bjorken sum rule (\ref{a3ga}) reflects the isospin SU(2) symmetry,
whereas the relation (\ref{3FD}) is a consequence of the SU(3) flavour 
symmetry treatment of the hyperon $\beta$-decays. While isospin symmetry 
is well established,  the growing precision of the measurements of magnetic 
moments and $G_A/G_V$ ratios in hyperon semi-leptonic decays \cite{PDG} may be indicating \cite{Gerasimov}-\cite{Ratcliffe} a non-negligible breakdown of SU(3) flavour symmetry and consequently of 
Eq. (\ref{3FD}). Several attempts have been already made \cite{Franklin, KL} 
to incorporate some symmetry breaking in the combined analysis of weak 
interaction data and polarized DIS data.\\

In this note we present a new study on the sensitivity of the polarized 
parton densities to  SU(3) breaking, using the world data on inclusive 
DIS \cite{data}.  
A significant improvement  of the previous results on this subject has been 
achieved. Firstly, a consistent NLO QCD treatment of the data has been carried out. Secondly, unlike the previous analyses, the effects of SU(3) symmetry breaking taken into account in our study are model-independent.
Further, we present results on the polarized parton densities 
{\it themselves}, not only on their first moments.  
Finally, we discuss the role of the different factorization schemes.\\

In the NLO QCD approximation the quark-parton decomposition of the 
spin-structure function $~g_1(x,Q^2)~$ has the following form:
\begin{eqnarray}
\nonumber
g_1(x,Q^2)&=&{1\over 2}\sum _{q} ^{N_f}e_{q}^2
[(\Delta q +\Delta\bar{q})\otimes (1 + {\alpha_s(Q^2)\over 2\pi}\delta C_q)\\
&+&{\alpha_s(Q^2)\over 2\pi}\Delta G\otimes \delta C_G],
\label{g1partons}
\end{eqnarray}
where $\Delta q(x,Q^2), \Delta\bar{q}(x,Q^2)$ and $\Delta G(x,Q^2)$ are
quark, anti-quark and gluon polarized densities which evolve in $Q^2$
according to the spin-dependent NLO DGLAP equations \cite{DGLAP}.
In (\ref {g1partons})
$\delta C_{q,G}$ are the NLO terms in the spin-dependent Wilson coefficient
functions and the symbol $\otimes$ denotes the usual convolution in Bjorken $x$ space. $\rm N_f$ is the number of flavours.

According to QCD, perfect data on the independent structure 
functions $g_1^p$ and $g_1^n(or g_1^d)$ {\it uniquely} determine the 
individual parton densities $(\Delta q +\Delta\bar{q})(x,Q^2)$ and 
$\Delta G(x,Q^2)$ in the DIS region.
It is the difference of their $Q^2$ evolution that enables them to be 
extracted from the data separately. However, bearing in mind the limited range in $x$ and $Q^2$ and the accuracy of the present data, it is immediately clear that the separation of the polarized parton densities from each other will not be very clear-cut. In other words, the unknown free parameters attached to the input parton densities at some arbitrary $Q^2=Q^2_0$ are correlated and not well determined from the fit to the inclusive DIS data alone. 
(Note that the attempts \cite{semiincl} to extract the polarized quark 
densities from {\it semi-inclusive} data 
are not entirely successful because, due to the quality of these data at present, many additional assumptions had to be made.)
That is why our knowledge on the first moments of the polarized parton 
densities coming from the hyperon semi-leptonic decays (Eqs. (\ref{a3ga}) and (\ref{3FD})) is usually used in addition to fix some of the free 
parameters.  It should be noted that the nonsinglet axial charges $a_3$ and
$a_8$ are $Q^2$ independent and this is the fact which helps us to link the 
information from both high and low-$Q^2$ regions.

We have performed an NLO QCD fit to the world data on $g_1^N(x,Q^2)$ 
\cite{data} using for the first moments of the quark densities the relations 
(\ref{a3ga}) and (\ref{3FD}). All details of our approach are given in 
\cite{spin98, JETscheme}. In contrast to our previous analyses
\cite{spin98}-\cite{newanal}, where we always used for $a_8$ its SU(3) 
symmetric value \cite{PDG96} $\rm 3F-D=0.579 \pm 0.025$, we incorporate now in our study the effects of the symmetry breaking. These effects have been taken into account in a model-independent way, i.e. we have used for $a_8$ different fixed values in the interval [0.36, 0.86]. These values cover the range obtained in various attempts to construct theorical models of SU(3) breaking.\\

The results of the analysis are presented in the JET scheme (see 
\cite{JETscheme} and references therein). The independence of the physical 
results on the choice of the factorization has been demonstrated in our 
previous papers \cite{JETscheme, newanal}. A remarkable property of the JET scheme is that the first moment of the singlet quark density,
$~\Delta \Sigma(Q^2)$, as well as $(\Delta s+\Delta\bar{s})(Q^2)$, 
the first moment of the strange sea quarks in the nucleon, are $Q^2$ 
independent quantities. Then, in the JET scheme is it meaningful to directly 
interpret $\Delta \Sigma$ as the contribution of the quark spins to the nucleon spin and to compare its value obtained from DIS region with the predictions of the different (constituent, chiral, etc.) quark models at low $Q^2$. 

It is useful to recall the transformation rules relating $\Delta \Sigma(Q^2)$ 
and $(\Delta s+\Delta\bar{s})(Q^2)$ in the JET and $\rm \overline{MS}$ schemes, the latter having been used in the previous papers \cite{KL} accounting for the SU(3) breaking effects:
\begin{equation}
\Delta \Sigma_{\rm JET}= \Delta \Sigma_{\rm \overline{MS}}(Q^2)
+N_f{\alpha_s(Q^2)\over 2\pi}\Delta G(Q^2)~,
\label{JETMS}
\end{equation}
\begin{equation}
(\Delta s+\Delta\bar{s})_{\rm JET}=
(\Delta s+\Delta\bar{s})_{\rm \overline{MS}}(Q^2) +
{\alpha_s(Q^2)\over 2\pi}\Delta G(Q^2)~,
\label{delstrrule}
\end{equation}
where $\Delta G(Q^2)$ is the first moment of the polarized gluon density
$\Delta G(x,Q^2)$ (note that $\Delta G$ is the same in the factorization 
schemes under consideration).

It is important to mention that the difference between the values of
$(\Delta s+\Delta\bar{s})$, obtained in the $\rm \overline{MS}$ and JET 
schemes could be large due to the axial anomaly. To illustrate how large it 
can be, we present the values of
$(\Delta s+\Delta\bar{s})$ at $Q^2=1~GeV^2$ obtained in our recent 
analysis \cite{newanal} of the world DIS data in the $\rm \overline{MS}$ 
and JET schemes (the SU(3) limit for $a_8$ has been used):
\begin{equation}
(\Delta s+\Delta\bar{s})_{\rm \overline{MS}}= -0.10 \pm 0.01,~~~~~
(\Delta s+\Delta\bar{s})_{\rm JET}= -0.06 \pm 0.01~.
\label{delsexp}
\end{equation}

Let us now comment briefly on how the deviation of $a_8$ from its SU(3) 
symmetric value influences the results on the polarized parton densities 
extracted from the DIS data. In order to demonstrate the sensitivity of the parton densities to the SU(3) breaking we present them (Figures 1-5) and their first moments (Table 1) for three typical values of $a_8$. All results are given at $Q^2=1~GeV^2~(\rm N_f=3$).  One can see from Table 1 that the values of $\chi^2$ are practically insensitive to the change of $a_8$, which means that any
of the solutions presented in this paper cannot be excluded by the present 
data. One can also conclude that except for the strange sea quarks and the 
gluons the other densities are essentially those determined by the SU(3) 
analysis of the data. It is important to stress that the singlet quark density, as well as its first moment, $\Delta \Sigma$ (the spin of the nucleon carried by the quarks), are virtually unchanged by the SU(3) breaking. 
The mean value of $\Delta \Sigma$ ranges from 0.34 to 0.40 and 
within the errors is not far from the value 0.60 expected in low-$Q^2$ quark 
models (see, e.g., \cite{lowQ2mod}).

\vskip 0.6 cm
\begin{center}
\begin{tabular}{cl}
&{\bf Table 1.} Sensitivity of the first moments of the polarized parton 
densities\\
&to SU(3) symmetry flavour symmetry breaking ($Q^2 = 1~GeV^2$, DOF=155).\\
&The SU(3) value 3F-D=0.58.
\end{tabular}
\vskip 0.6 cm
\begin{tabular}{|c|c|c|c|c|c|c|} \hline
 ~~$a_8$~~&$\chi^2$&$\Delta u + \Delta\bar{u}$ &
 $-(\Delta d + \Delta\bar{d})$ & $-(\Delta s + \Delta\bar{s})$ 
&$\Delta \Sigma$&$\Delta G$\\ \hline
 0.40 & 127.4 &0.81~$\pm$~0.02 & 0.45~$\pm$~0.02 &
 0.02~$\pm$~0.01 & 0.34~$\pm$~0.05 &0.13~$\pm$~0.14\\
 3F-D& 128.3 & 0.86~$\pm$~0.02 & 0.40~$\pm$~0.02 & 0.06~$\pm$~0.01 &
 0.40~$\pm$~0.04 & 0.57~$\pm$~0.14\\
 0.86 & 127.4 & 0.90~$\pm$~0.02 & 0.35~$\pm$~0.02 & 0.15~$\pm$~0.02 &
 0.40~$\pm$~0.06 & 0.84~$\pm$~0.30\\ \hline
\end{tabular}
\end{center} 
\vskip 0.6 cm

Contrary to the singlet and non-strange quarks, the strange sea 
polarization changes significantly when flavour SU(3) symmetry is broken
(see Fig. 3). Comparing with the SU(3) case the strange sea contribution to 
the nucleon spin is reduced by a factor of three when $a_8=0.40$ and enhanced 
by a factor of two and a half when $a_8=0.86$. In the case $a_8=0.40$ the 
strange polarization is consistent with zero in agreement with the usual 
assumption in low-$Q^2$ quark models. This fact, contrary to what is 
sometimes claimed (see, e.g.,\cite{RatST}), does not help to solve the
"spin crisis in the naive parton model" since the value of $\Delta \Sigma$ remains virtually unchanged.
The mean value of the gluon polarization $\Delta G$ ranges from 0.13 to 0.84, 
but because of the large errors these values are consistent within two 
standard deviations. As seen from Table 1, although a significant 
improvement of the quality of the data since the EMC experiment has been 
achieved, the present data still do not exclude a vanishing gluon polarization 
in the DIS region if the value of $a_8$ is considerably smaller than the 
SU(3) value.

It is important to emphasize that although the axial charge $a_8$ and the 
strange sea quarks $(\Delta s + \Delta\bar{s})$ cannot be well separated
using the current DIS data alone, $a_8$ and $(\Delta s + \Delta\bar{s})$ 
are {\it independent} quantities ($\Delta q_8(x,Q^2)$ and 
$(\Delta s + \Delta\bar{s})(x,Q^2)$ evolve with $Q^2$ in different ways). 
That is why any combined analysis of the DIS and the hyperon $\beta$-decays data, in which the issue of the SU(3) breaking is model-dependent, has to take account of this fact. Otherwise, such an analysis is inconsistent, as was already discussed in detail in our note \cite{hypDIS}. \\

In conclusion, we note that a further improvement of the situation is expected to come from: {\it i)} the current KTeV experiment at Fermilab on the $\Xi^0~\beta$-decay,
$\Xi^0 \rightarrow \Sigma^{+}e\bar{\nu}$, (see \cite{RatST} and references
therein) and {\it ii)} a combined analysis of inclusive and semi-inclusive 
present and future DIS data. While the KTeV experiment will help in 
clarifying the issue of the SU(3) breaking, and thereby to find the proper 
value of $a_8$, the future DIS experiments will be very important for a precise determination of the polarized parton densities (and, in particular, $a_8$) {\it independently} of the information coming from the low-$Q^2$ region. The consistency of the values of $a_8$ obtained from the high and low-$Q^2$ regions will be a good test for our understanding of the spin properties of the nucleon.\\
\vskip 3mm
This research was partly supported by a UK Royal Society Collaborative
Grant, by the Russian Foundation for Basic Research, Grant No 
00-02-16696 and by the Bulgarian National Science Foundation. \\

\newpage
\noindent
{\bf Figure Captions}
\vskip 3mm
\noindent 
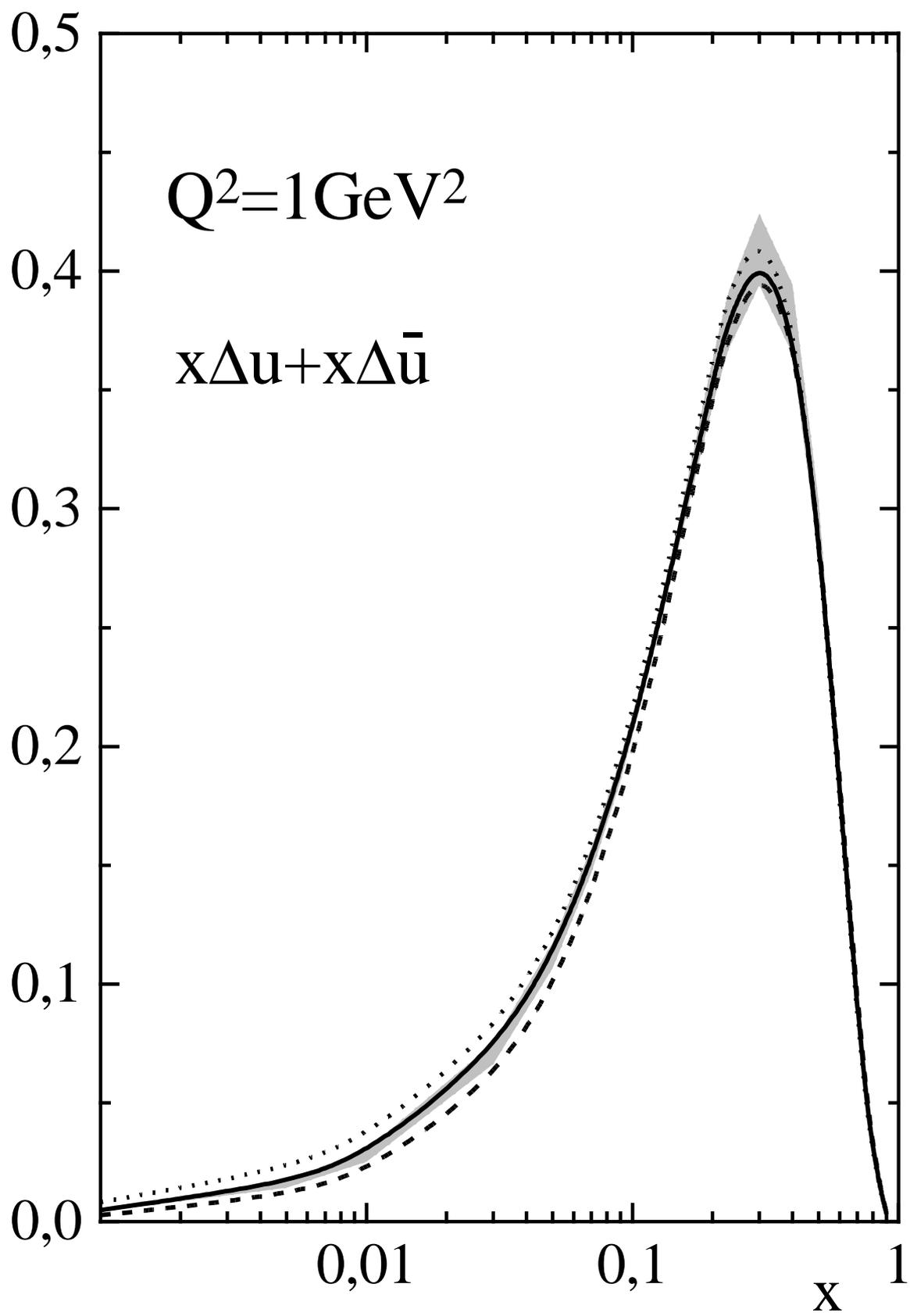
{\bf Fig. 1.} Next-to-leading order polarized parton density 
$x(\Delta u + \Delta\bar{u})$ at $~Q^2=1~GeV^2~$ (JET scheme). The solid, 
short-dashed and long-dashed curves correspond to the values 0.58 (SU(3) limit), 0.86 and 0.40 for the axial charge $a_8$, respectively. The error band accounts for the statistical and systematic uncertainties.\\

\noindent
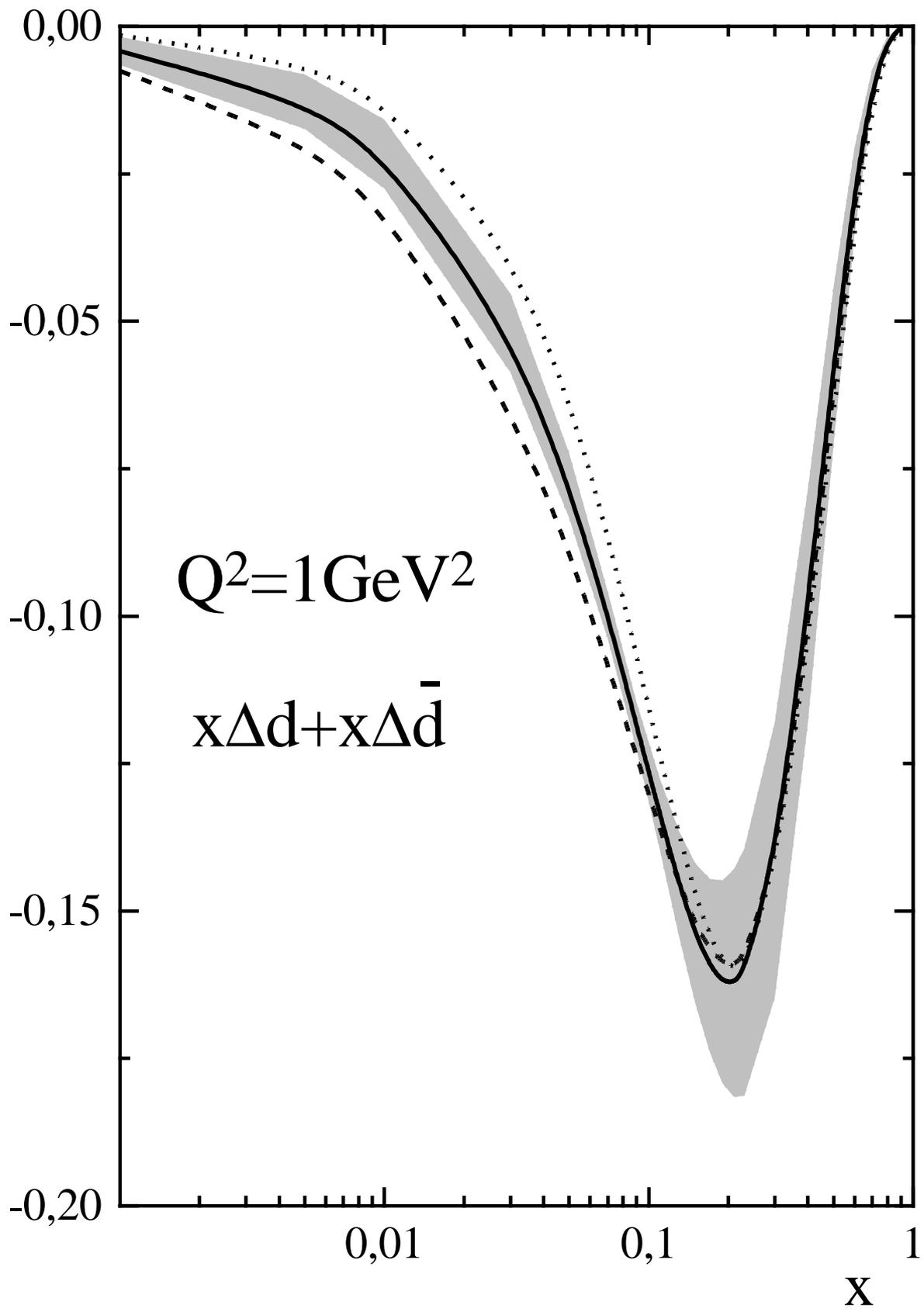
{\bf Fig. 2.} Next-to-leading order polarized parton density 
$x(\Delta d + \Delta\bar{d})$ at $~Q^2=1~GeV^2~$ (JET scheme). 
See caption to Fig. 1.\\

\noindent
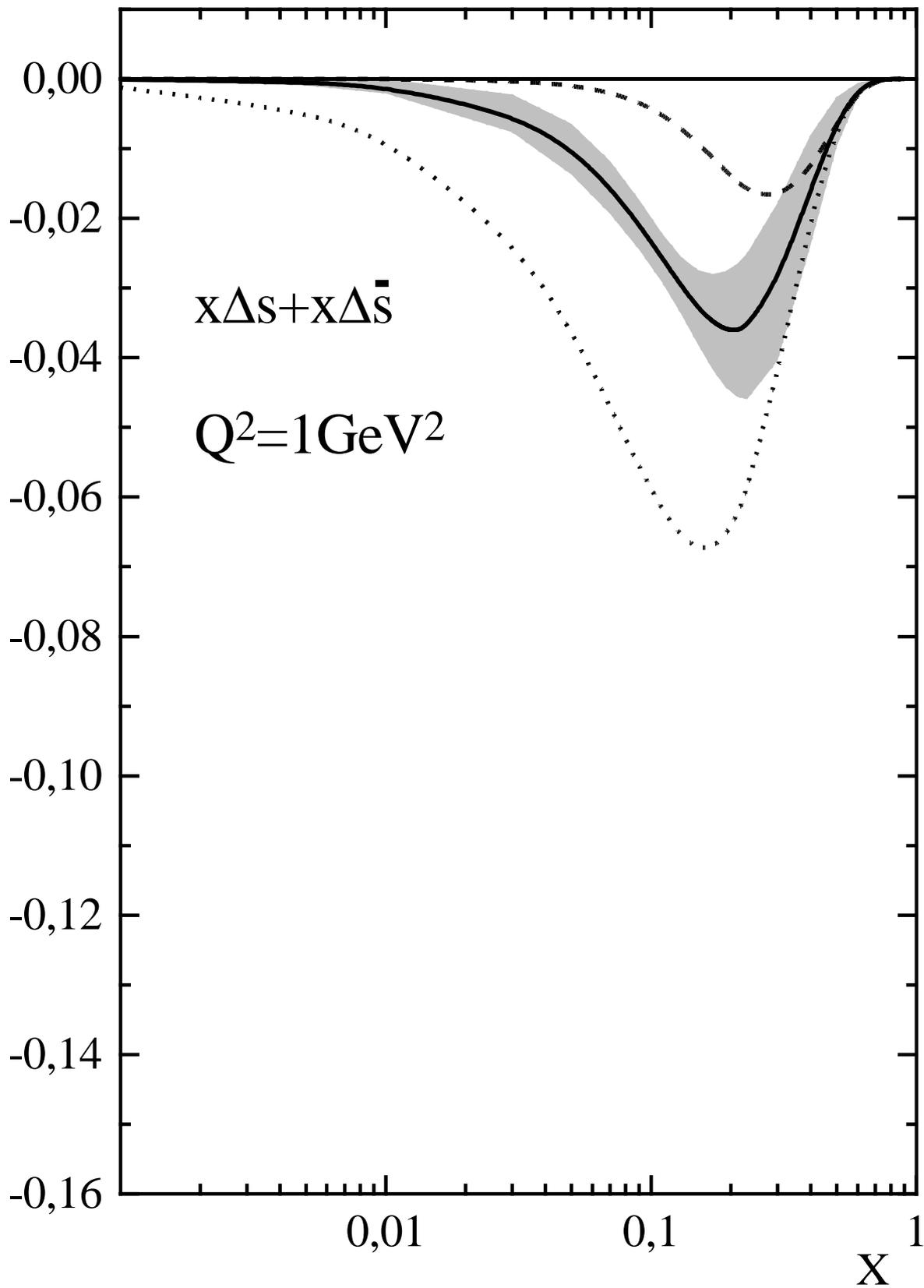
{\bf Fig. 3.} Next-to-leading order polarized strange sea quark density 
$x(\Delta s + \Delta\bar{s})$ at $~Q^2=1~GeV^2~$ (JET scheme). 
See caption to Fig. 1.\\

\noindent
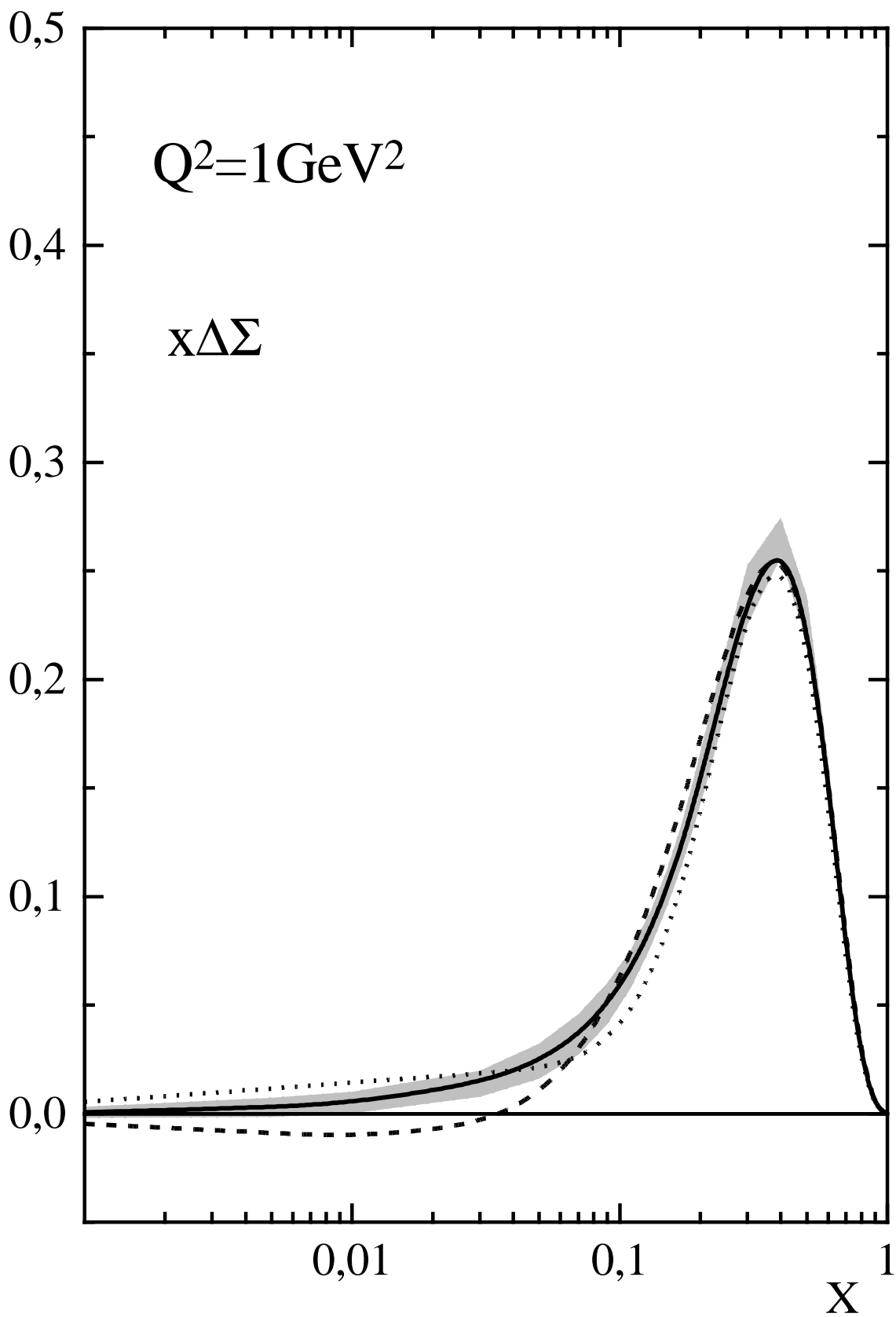
{\bf Fig. 4.} Next-to-leading order polarized singlet quark density
$x\Delta \Sigma$ at $~Q^2=1~GeV^2~$ (JET scheme). 
See caption to Fig. 1.\\

\noindent
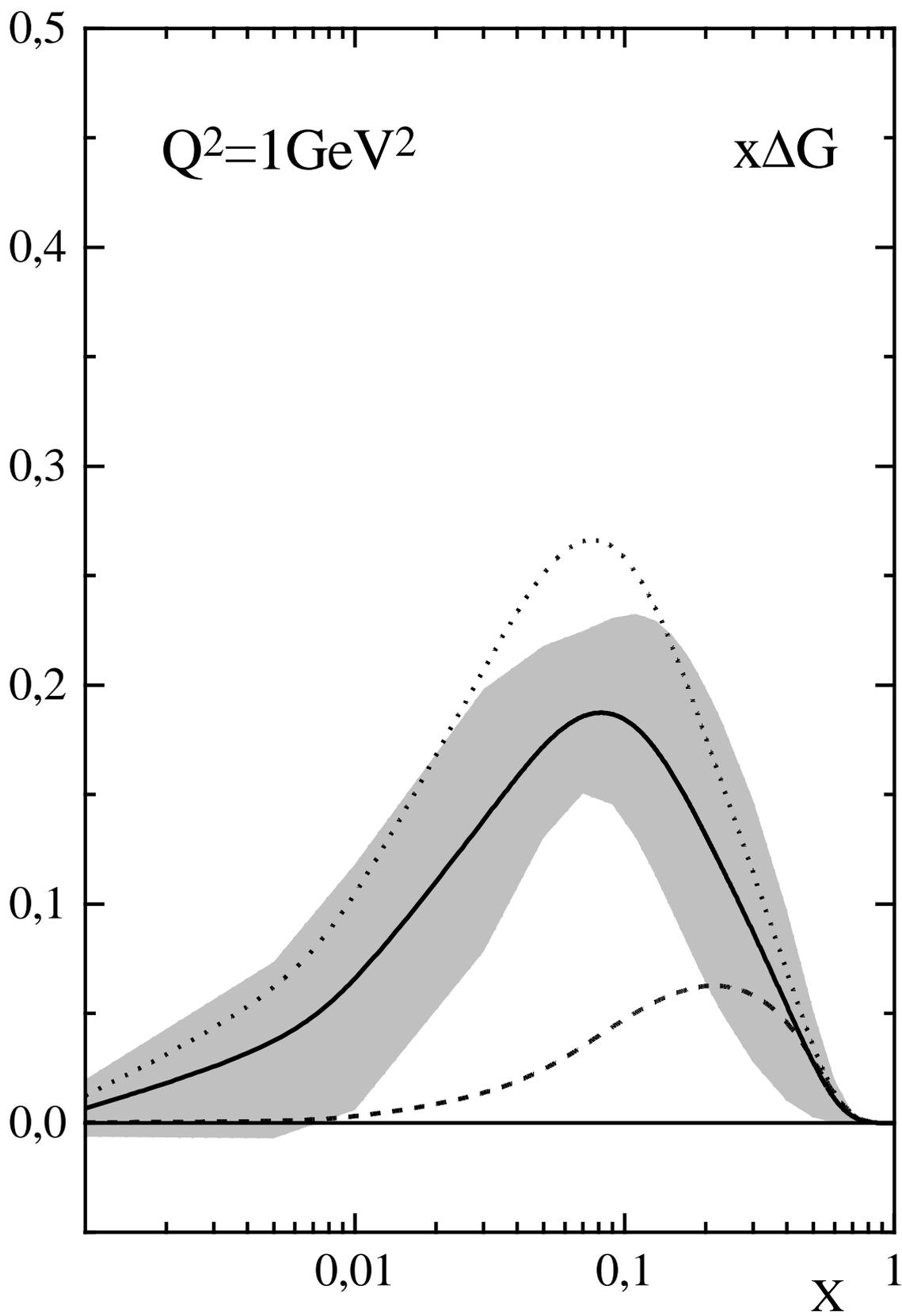
{\bf Fig. 5.} Next-to-leading order polarized  gluon density 
$x\Delta G$ at $~Q^2=1~GeV^2~$ (JET scheme). 
See caption to Fig. 1.\\

\end{document}